\documentclass[useAMS, usenatbib, a4paper]{mn2e}
\usepackage{aas_macro, amsmath, amssymb, color, graphicx, multirow, setspace}
\begin{document}

\title[Satellites of LMC-Mass Dwarfs]{Satellites of LMC-Mass Dwarfs: Close Friendships Ruined by Milky Way Mass Halos}

\author[A. J. Deason et al.]{A. J. Deason$^{1}$\thanks{E-mail: alis@ucolick.org}\thanks{Hubble Fellow}, A. R. Wetzel$^{2,3}$, S. Garrison-Kimmel$^{4}$, V. Belokurov$^{5}$\\
$^{1}${Department of Astronomy and Astrophysics, University of California Santa Cruz, Santa Cruz, CA, USA}\\
$^{2}${TAPIR, California Institute of Technology, Pasadena, CA, USA}\\
$^{3}${Observatories of the Carnegie Institution for Science, Pasadena, CA, USA}\\
$^{4}${Center for Cosmology, Department of Physics and Astronomy, University of California, Irvine, CA, USA}\\
$^{5}${Institute of Astronomy, University of Cambridge, Madingley Road, Cambridge, CB3 0HA, UK}}

\date{\today}

\pagerange{\pageref{firstpage}--\pageref{lastpage}} \pubyear{2015}

\maketitle

\label{firstpage}

\begin{abstract}
Motivated by the recent discovery of several dwarfs near the Large Magellanic Cloud (LMC), we study the accretion of massive satellites onto Milky Way (MW)/M31-like halos using the ELVIS suite of N-body simulations. We identify 25 surviving LMC-mass subhalos, and investigate the lower-mass satellites that were associated with these subhalos before they fell into the MW/M31 halos.  Typically, 7\% of the overall $z=0$ satellite population of MW/M31 halos were in a surviving LMC-group before falling into the MW/M31 halo. This fraction can vary between 1\% and 25\%, being higher for groups with higher mass and/or more recent infall times. Groups of satellites disperse rapidly in phase space after infall, and their distances and velocities relative to the group center become statistically similar to the overall satellite population after $4-8$ Gyr. We quantify the likelihood that satellites were associated with an LMC-mass group as a function of both distance and velocity relative to the LMC at $z=0$. The close proximity in distance of the nine Dark Energy Survey candidate dwarf galaxies to the LMC suggest that $\sim2-4$ are likely associated with the LMC. Furthermore, if several of these dwarfs are genuine members, then the LMC-group probably fell into the MW very recently, $\lesssim2$ Gyr ago. If the connection with the LMC is established with follow-up velocity measurements, these ``satellites of satellites'' represent prime candidates to study the affects of group pre-processing on lower mass dwarfs.

\end{abstract}

\begin{keywords}
Galaxy: formation --- Galaxy: halo --- galaxies: dwarf --- galaxies: Magellanic Clouds
\end{keywords}

\section{Introduction}
The $\Lambda$ cold dark matter ($\Lambda$CDM) model predicts an
abundance of substructure on all (observable) mass scales. Hundreds of subhalos are predicted to surround
Milky Way (MW) mass halos (\citealt{klypin99}), which can be likened to a scaled down
version of the thousands of substructures associated with
clusters (\citealt{moore99}). This trend, presumably, continues to smaller mass scales,
whereby dwarf galaxies can also host several substructures. Recent
discoveries of dwarf-dwarf accretion \citep{delgado12, belokurov13}
present tantalizing observational evidence for the existence of
such ``sub-structure of sub-structure''.

Despite the hierarchical nature of dark matter halos, we generally ignore the possibility that some of the MW satellites may have been part of a group of subhalos before they fell into the Galaxy. The relatively
unexplored population of sub-subhalos or ``satellites of satellites''
is strongly linked to the most massive structures in the MW halo, as
these are the potential vehicles that dragged in several low
mass dwarfs. For example, \cite{wetzel15} showed that a significant
fraction ($\sim 30$ \%) of low-mass subhalos ($M_{\rm star} \lesssim
10^5M_\odot$) likely fell into a MW-type host as a satellite of a more
massive subhalo, and $>50\%$ were in a group before infall. The most likely culprit in our own Galaxy is the
Large Magellanic Cloud (LMC). This massive dwarf already has one
obvious companion, the Small Magellanic Cloud (SMC), but it likely had
several other companions in the past.

Numerous works have attempted to connect the LMC to other known dwarfs
in the MW. \cite{lyndenbell76} first suggested the idea of a ``Greater
Magellanic Galaxy'', and he later postulated the association of
several of the classical dwarfs with the Magellanic complex
(\citealt{lyndenbell82}; \citealt{lyndenbell95}). More recently,
\cite{donghia08} suggested that seven of the MW satellites could have
been part of a late infalling LMC group. In contrast, \cite{sales11}
use an LMC-analog ``case-study'' in a cosmological simulation to show that most of
the classical dwarfs show little evidence for an association with the
LMC.   However, the authors do prophetically suggest that faint, previously unnoticed MW satellites could be lurking in the vicinity of the Clouds.

The discovery of very low luminosity galaxies ($L \lesssim 10^5
L_\odot$) in the MW (e.g, \citealt{willman05}; \citealt{belokurov06,
  belokurov07}) has, until recently, been restricted to the Sloan
Digital Sky Survey (SDSS) footprint, as most of the ``ultra-faint''
dwarf population have been discovered using SDSS imaging. However,
uncharted territory beneath declination $\delta=-30^\circ$ has recently been explored with the first data release of the Dark Energy
Survey (DES). Two independent groups (\citealt{bechtol15};
\citealt{koposov15}) unveiled eight and nine candidate dwarf galaxies
correspondingly in the DES data. Curiously, these satellites are
mostly of the ``ultra-faint'' variety and are in close proximity to
the LMC.

In \cite{wetzel15} we showed that most of the past satellites of an
LMC-mass dwarf are likely lower mass subhalos, likened to the
ultra-faints. Thus, the finding of several low luminosity dwarfs in
close proximity to the LMC could potentially confirm a generic prediction of hierarchical structure formation. In this work, we use cosmological simulations to study the satellite populations of LMC-mass dwarfs accreted onto
MW/M31-mass halos, in order to understand the potential association
between the newly discovered DES satellites and the LMC in a
cosmological context.

\section{Numerical Methods}

\subsection{ELVIS Simulations}

To study the satellite populations of LMC-mass dwarfs, we use ELVIS
(Exploring the Local Volume in Simulations, \citealt{gk14}), a suite of 48 high-resolution, zoom-in simulations of Milky Way/M31 mass halos ($M_{\rm vir} = 1 - 3 \times 10 ^ {12} M_\odot$). Half of the ELVIS halos reside in a paired configuration with separations and relative velocities similar to those of the MW-M31 pair, while the remainder are highly isolated halos mass-matched to those in the pairs.

Within the high-resolution, zoom-in volumes (spanning 2-5 Mpc in size), the particle mass is $1.9 \times 10 ^ 5 M_\odot$ and the Plummer-equivalent force softening is 140 pc. ELVIS adopts a cosmological model based on \textit{Wilkinson
  Microwave Anisotropy Probe} WMAP7 \citep{larson11}, with the following $\Lambda$CDM parameters: $\sigma_8=0.801$, $\Omega_M = 0.266$, $\Omega_\Lambda = 0.734$, $n_s = 0.963$ and $h = 0.71$. See \citet{gk14} for more details on ELVIS.

\subsection{Finding and tracking subhalos}

Dark matter subhalos are identified in ELVIS using the six-dimensional halo finder \textsc{rockstar} \citep{behroozi13a}, and merger trees are constructed using the \textsc{consistent-trees} algorithm \citep{behroozi13b}.  

For each subhalo, we assign its primary progenitor (main branch) as
the progenitor that contains the largest total mass summed from the
subhalo masses over all preceding snapshots in that branch. We then
compute the maximum (peak) mass, $M_{\rm peak}$, ever reached by the
main branch of a progenitor.

Throughout this work, we only consider subhalos with $M_{\rm peak} >
10^8 M_{\odot}$ (or $M_{\rm star} \gtrsim 5 \times 10^2 M_\odot$); \cite{gk14} found that subhalos down to this mass-threshold do not suffer from resolution and numerical disruption issues.

\begin{figure}
  \centering
   \includegraphics[width=8.0cm, height=6.4cm]{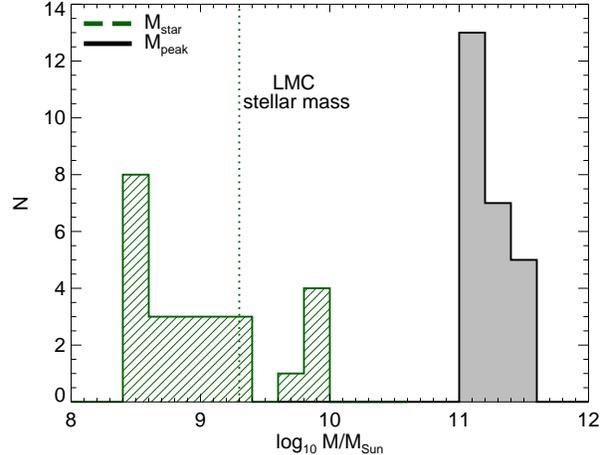}
   \caption{\small Distribution of peak subhalo mass, $M_{\rm peak}$ (solid gray), and stellar mass, $M_{\rm star}$ (hashed green), for the 25 satellites with $M_{\rm peak} > 10^{11}M_\odot$ (masses near that expected for the LMC) in the MW/M31 hosts at $z=0$ in the ELVIS simulation suite. These are the (surviving) ``LMCs'' that we will discuss in the remainder of the paper. The dotted line indicates the observed stellar mass of the LMC.}
\label{fig:mlmc}
\end{figure}

\subsection{Sample of LMC-mass satellites}
We select a sample of LMC-mass satellites of MW/M31 hosts at $z=0$
using all 48 (paired and isolated)\footnote{We find no significant
  differences in our results when just the paired halos are used.}
halos in the ELVIS simulation suite. We select $z=0$ satellites with
$M_{\rm peak} > 10^{11}M_\odot$ ($M_{\rm star} \gtrsim 3 \times 10^{8}M_\odot$). This lower mass cut is approximately a factor of two lower than the LMC mass ($M_{\rm peak} \sim 2 \times 10^{11}$ for $M_{\rm star} \approx 2 \times 10^{9}$, \citealt{vdm02}). Stellar masses are estimated for the
dark matter subhalos using the $M_{\rm star}-M_{\rm peak}$ relation
derived in \cite{gk14}. We exclude the satellite in the
\textit{Sonny \& Cher} paired simulation that has a mass comparable to its host
halo ($M_{\rm peak} \sim 7 \times 10^{11} M_\odot$).

A significant number of the host halos in ELVIS ($\sim 50$\%) do not have
\textit{any} satellites more massive than $M_{\rm peak} > 10^{11}
M_\odot$, while some halos have more than one LMC satellite. It is
unlikely that relatively low mass MW/M31 halos with $M_{\rm vir}
\approx 10^{12}M_\odot$ host very massive satellites (e.g. \citealt{boylan10}), so we are biased towards the more massive
host halos in the ELVIS suite (typically $\langle M_{\rm vir} \rangle \sim 2 \times 10^{12}M_\odot$). We also note that some of the paired halos
were selected to have a satellite companion with mass similar to the
LMC ($M_{\rm star} \sim 10^9M_\odot$, see \citealt{gk14}). Thus, our
sample is not an unbiased (random) selection of MW/M31 mass hosts.

Our final sample comprises 25 LMC-mass dwarf satellites at
$z=0$. Fig. \ref{fig:mlmc} shows the distribution of their stellar and
peak dark matter masses.

\subsection{Finding the (surviving) satellites of LMC dwarfs \textit{prior} to infall onto the MW/M31 host}

\begin{figure}
  \centering
   \includegraphics[width=8.5cm, height=5.67cm]{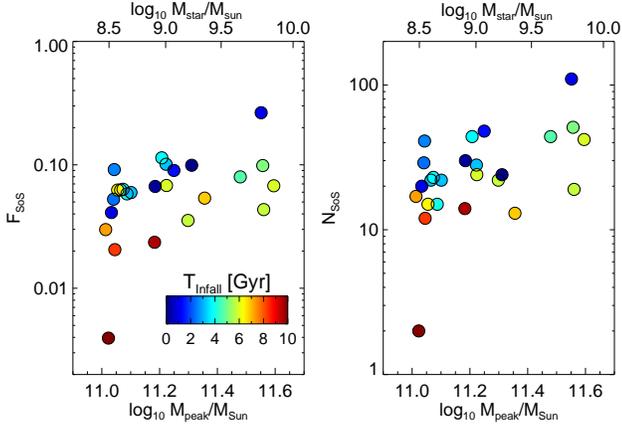}
   \caption{\small The fraction (left panel) and number (right panel) of satellites of MW/M31 hosts at $z=0$ that were satellites of the surviving LMC dwarfs prior to its infall onto the MW/M31 host. The peak mass (stellar mass) of the LMCs is shown on the bottom (top) x-axis. The color scheme indicates the time since infall of the accretion events. Recent and/or massive accretion events contribute significant numbers of ``satellites of satellites'' to the $z=0$ satellite population.}
\vspace{10pt}
\label{fig:fsos}
\end{figure}

We trace back all $z=0$ satellites of MW/M31 hosts and identify those
that were satellites of a surviving LMC dwarf anytime \textit{before} infall
onto the MW/M31 hosts. We assume \textit{all} subhalos with $M_{\rm peak} > 10^8M_\odot$ host luminous galaxies.  We impose that a subhalo must remain a
satellite for at least two consecutive time-steps ($\Delta T \approx 400$ Myr) in the simulations
to avoid counting particularly transient (and likely non-meaningful)
crossings just within $R_{\rm vir}$.

Note that the LMC dwarfs themselves are now also satellites of
MW/M31 hosts, but prior to infall are the group centrals. 

In total, we identify $N=734$ MW/M31 satellites today that were once
satellites of LMC dwarfs, where these ``LMCs'' are still intact today. These ``satellites of satellites'' comprise approximately 7\% of the surviving satellite population of
MW/M31 hosts at $z=0$, and have typical stellar masses of $M_{\rm star} =10^{3}-10^{5}M_\odot$ (comparable to the ultra-faint dwarf
galaxy population). This fraction is lower than in \cite{wetzel15}
because we only consider the subset of satellites that were
satellites of a \textit{surviving}\footnote{\cite{wetzel15} show that approximately half of the overall population of group centrals have merged/disrupted by $z=0$.} LMC satellite before infall. In this work, we only consider subhalos within the MW/M31 hosts today, and do not include ``field'' subhalos (i.e. outside of $R_{\rm vir}$ today) that could have been associated with an LMC-mass dwarf in the past. It is worth noting, however, that these associations do exist, and this could be an interesting population to study in future work.

\section{Results}
\subsection{Satellites of LMC-mass dwarfs}

\begin{figure*}
  \centering
   \includegraphics[width=16cm, height=6.4cm]{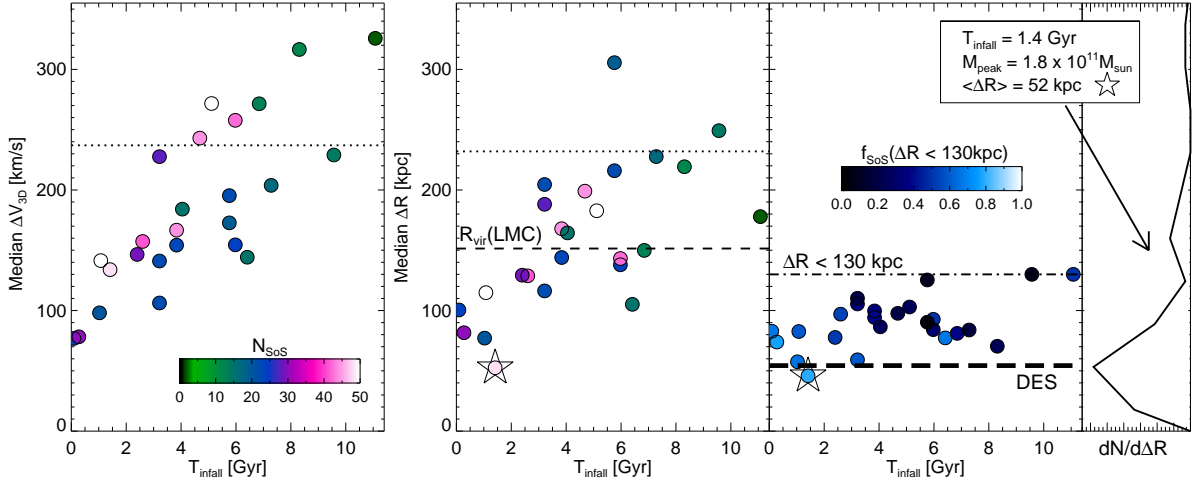}
   \caption{\small The median differences in 3D velocity (left panel)
     and 3D distance (right panels) at $z=0$ between the
     LMC dwarfs and the satellites that were associated with
     them in a group before falling into the MW/M31 hosts. LMCs that fell in at early times have the largest
     differences in phase space with their satellites today. The
     colors indicate the number of surviving group members. The black
     dotted lines indicate the average velocity/radial difference
     between \textit{all} satellites in MW/M31 hosts and the LMC satellite at $z=0$. We also show the approximate virial radius for an LMC-mass subhalo with the short-dashed line. The middle right panel shows the median difference in
     configuration space for satellites with $\Delta R < 130$
     kpc. This is a rough estimate for the maximum $\Delta R$ probed
     by the DES survey around the LMC. The colors indicate the fraction of surviving group members that have $\Delta R < 130$ kpc. The black dashed line shows the
     median distance between the DES dwarfs and the LMC.  The furthest right panel  shows the distribution of $\Delta R$ for one massive group
     (indicated by the star symbol) with low median $\Delta R$.}
\label{fig:delta_vr}
\end{figure*}

Fig. \ref{fig:fsos} shows the fraction (left panel) and number (right
panel) of $z=0$ satellites that were associated with a surviving LMC dwarf
before infall onto the MW/M31, against the mass of the group
central. The color scheme indicates the infall time\footnote{Throughout we use ``infall time'' to define the time \textit{since} infall of a subhalo onto a host halo.} of the accretion
events onto the MW/M31 hosts (blue=recent infall, red=early infall).

Unsurprisingly, more massive dwarfs have more abundant satellite
populations. There is also a dependence on infall time
onto the MW/M31 host. At a given mass, groups accreted more recently
have more surviving members at $z=0$. 

\subsection{Phase-space associations at $z=0$}

\begin{figure}
  \centering
   \includegraphics[width=8.5cm, height=8.5cm]{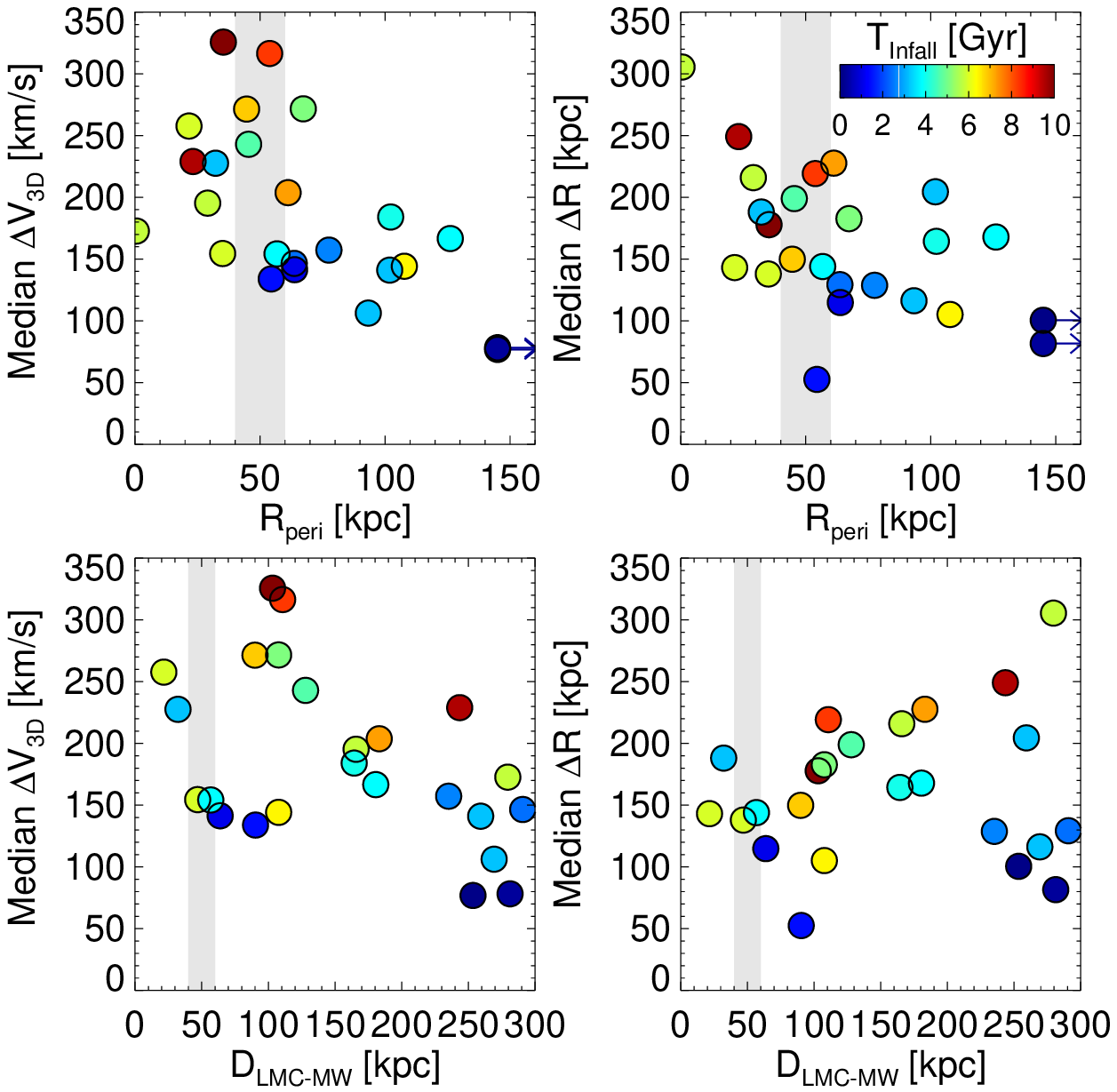}
   \caption{\small The median differences in 3D velocity (left panels) and 3D distance (right panels) at $z=0$ between the LMC dwarfs and the satellites that were associated with them in a group before falling into the MW/M31 hosts, as a function of the distance of closest approach to the MW/M31 hosts ($R_{\rm peri}$, top panels) and the $z=0$ distance between the LMC-mass dwarf and the MW/M31 ($D_{\rm LMC-MW}$, bottom panels). LMCs with larger pericenters tend to have smaller differences in phase space with their satellites today. However, there is a lot of scatter in this relation, particularly for dwarfs with $R_{\rm peri} \lesssim 100$ kpc. For example, dwarfs with $40 <R_{\rm peri}/\mathrm{kpc} < 60$ (the approximate pericenter of the LMC, illustrated with the gray bands) can have a wide range of infall times and phase-space differences. Note that two dwarfs with $R_{\rm peri} > 150$ kpc are shown as lower limits in these plots. The trends with present day distance ($D_{\rm LMC-MW}$, bottom panels) are weaker. The colors indicate the infall times of the LMC-mass dwarfs. }
\label{fig:rperi}
\end{figure}

We now consider the current ($z=0$) association in phase-space between
the LMC dwarfs and their former satellite population. In
Fig. \ref{fig:delta_vr} we show the median velocity (left panel) and 3D distance (right panels) between the ``LMCs'' and their
past group members as a function of infall time onto the MW/M31 hosts. We note that here, and throughout this work, we use infall time as an orbital constraint in our analysis. We find that infall time, rather than either the current distance from the MW/M31 host or orbital pericentric distance, shows the strongest relation with the dispersal of groups in phase-space (see below and Fig. \ref{fig:rperi}).

After infall, groups become more dispersed in phase-space over time
(see also \citealt{sales11}).  For comparison, we show the typical
average velocity/distance difference between \textit{all} satellites
of MW/M31 hosts at $z=0$ and the group centrals with the dotted
lines. Groups accreted more than $\sim 5-6$ Gyr ago are well mixed in
phase-space today. For illustration, the far-right panel of Fig. \ref{fig:delta_vr} shows the distribution of $\Delta R$ for one LMC-group with low median $\Delta R$. Note that this group also has similar dynamical properties to the observed LMC-system (see Section \ref{sec:dyn}).

In the middle right panel, we show the median difference in configuration
space for satellites with $\Delta R < 130$ kpc from the group
central. This is a rough estimate for the maximum $\Delta R$ probed by
the DES survey around the LMC (see \citealt{koposov15} Fig. 20). The proximity of the DES satellites to the LMC is striking, especially compared to the general population of group members in the simulations. This proximity in configuration space not only suggests a likely association between the DES dwarfs and the LMC, but, if several of these dwarfs are genuine group members, then it implies a very recent infall time for the LMC-group. Note that the most recent
observational constraints on the orbits of the LMC/SMC suggest a
recent infall time for this group (see e.g, \citealt{besla07};
\citealt{boylan11}; \citealt{rocha12}; \citealt{kalli13}).

Qualitatively, a picture similar to the above has been painted by the
study of \citet{sales11}. However, here we present the first {\it
  quantitative} evidence of a pronounced correlation between $z=0$
scatter in the phase-space exhibited by the ``satellites of the
satellites'' for a statistically significant sample of accretion
configurations. 

In Fig. \ref{fig:rperi} we show the median velocity and 3D distance at $z=0$ between the ``LMCs'' and the MW/M31 satellites that were associated with them prior to infall, as a function of the pericenter distance (top panels) and the $z=0$ distance (bottom panels) from the MW/M31 hosts. Groups with smaller pericenters show a larger dispersal in phase-space, but the scatter between systems is large, particularly $R_{\rm peri} \lesssim 100$ kpc. For pericenters similar to the observed closest approach of the LMC ($\sim 50$ kpc), groups can be highly dispersed or can remain in close proximity at $z=0$ (e.g. Median $\Delta R \sim 50-230$ kpc). We also note that LMC-mass dwarfs with pericenters close to the observed value of the LMC ($\sim 50$ kpc), can have a wide range of infall times ($\sim 1-9$ Gyr). The trends with present day distance from the MW/M31 hosts are also relatively weak, particularly for dwarfs with $D_{\rm LMC-MW} \lesssim 150$ kpc . Finally, we also find that the dispersal in phase-space of groups, at least for our sample, is only weakly dependent on the orbital eccentricity of the LMC-mass dwarfs.

\subsection{Dynamical LMC-analogues}
\label{sec:dyn}

\begin{table}
\centering
\renewcommand{\tabcolsep}{0.1cm}
\renewcommand{\arraystretch}{0.6}
\begin{tabular}{|c c c c c c |}\hline
$M_{\rm LMC}$ & $R_{\rm peri}$ & $V_{R}(R_{\rm peri})$ & $V_{T}(R_{\rm peri})$ & $T_{\rm infall}$ &  $N_{\rm SoS}$ \\
$[M_\odot \times 10^{11}]$& [kpc] & [km s$^{-1}$] & [km s$^{-1}$] & [Gyr] & \\
\hline
1.1 & 78 & 78 & 327 & 2.6 & 41\\
1.0  & 61 & 45 & 313 & 7.3 & 17\\
1.8  & 54 & 106 & 341 & 1.4 & 48 \\
\hline
 \end{tabular}
\caption[]{Sample of 3 LMC-mass dwarfs with dynamical properties similar to the observed LMC system (see main text for details).}
\label{tab:analogues}
\end{table}
\begin{figure}
  \centering
   \includegraphics[width=8.5cm, height=6.4cm]{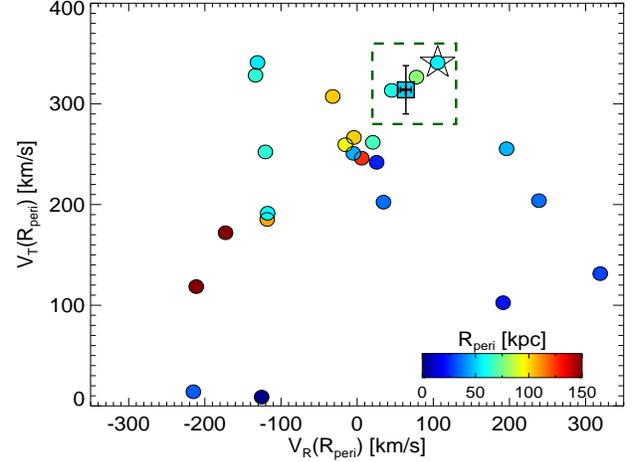}
   \caption{\small The tangential and radial velocity of our sample of LMC-mass dwarfs at pericenter. The points are colored by pericentric distance. The square symbol (and error bar) indicates the observed velocity components of the LMC (\citealt{kalli13}). The dashed-green box indicates 3 dwarfs in our sample with similar dynamical properties to the observed LMC. The star symbol indicates the dwarf highlighted in Fig. \ref{fig:delta_vr} with a very low median $\Delta R$.}
\label{fig:dyn}
\end{figure}

Our sample of LMC-mass dwarfs is based purely on peak subhalo mass (see Fig. \ref{fig:mlmc}). We now select a subsample of these dwarfs based on observational dynamical constraints. Figure \ref{fig:dyn} shows the radial and tangential motion of our sample of 25 LMC-mass dwarfs at pericenter, where the points are colored by pericentric distance. The estimated motion of the LMC derived by \cite{kalli13} is shown with the star symbol on this plot ($V_{\rm R, LMC}=64 \pm 7$ km s$^{-1}$, $V_{\rm T, LMC}=314 \pm 24$ km s$^{-1}$, $R_{\rm peri}=50$ kpc). The dashed green box highlights 3 dwarfs with similar velocity and pericentric distance to that observed for the LMC, and the properties of these dwarfs are listed in Table \ref{tab:analogues}. Note that one of these dwarfs was highlighted in Fig. \ref{fig:delta_vr} and has a very low median $\Delta R$.

In the following subsection, we investigate the probability of group membership based on the phase-space dispersal of group members at $z=0$. Here, we use both the full sample of 25 LMC-mass dwarfs and the subsample of 3 dwarfs that we label as ``dynamical analogues''.

\subsection{Likelihood of group-membership}

\begin{figure*}
  \centering
   \includegraphics[width=16cm, height=5.33cm]{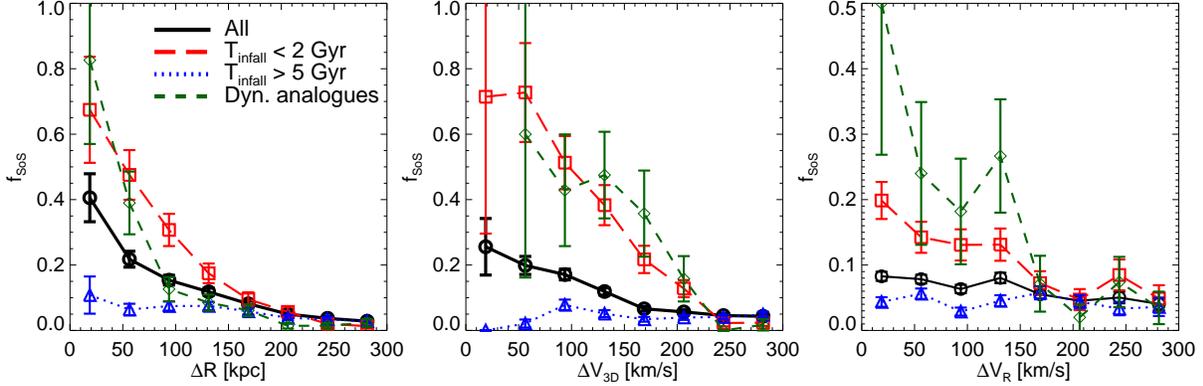}
   \caption{\small The fraction of all MW/M31 satellites at $z=0$ that were a satellite of a surviving LMC-mass dwarf before infall onto the MW/M31 host as a function of 3D distance (left panel) and velocity (middle and right panels) difference from the LMC today. The long-dashed red and dotted blue lines are for groups accreted recently ($T_{\rm infall} < 2$ Gyr) and early ($T_{\rm infall} > 5$ Gyr), respectively. Only groups accreted recently show a close-proximity in phase-space at $z=0$. The dashed-green lines are for the dynamical LMC-analogues (see Section \ref{sec:dyn}).}
\label{fig:delta_prob}
\end{figure*}

Fig. \ref{fig:delta_prob} presents the probability of a past
association with an LMC-mass dwarf as a function of distance (left
panel) and velocity (middle and right panels) from the massive group
central. This now includes ``interloping'' satellites near the LMC at $z = 0$ that were not satellites of the LMC-mass host prior to MW/M31 infall. Dwarfs more closely related in configuration or velocity space are more likely to have been group members before infall. For example, $> 25\%$ of dwarfs within 50 kpc of an LMC-mass dwarf were likely associated with this dwarf before infall. We show the 3D velocity difference and radial velocity difference ($\Delta V_R=|V_{R,\mathrm{SoS}}-V_{R,\mathrm{LMC}}|$) between group members in the middle and right-hand panels, respectively. Although, 3D velocity information gives a cleaner distinction between members and non-members, radial velocities can also be used to assign membership probabilities.

\begin{figure}
  \centering
   \includegraphics[width=8.5cm, height=12.75cm]{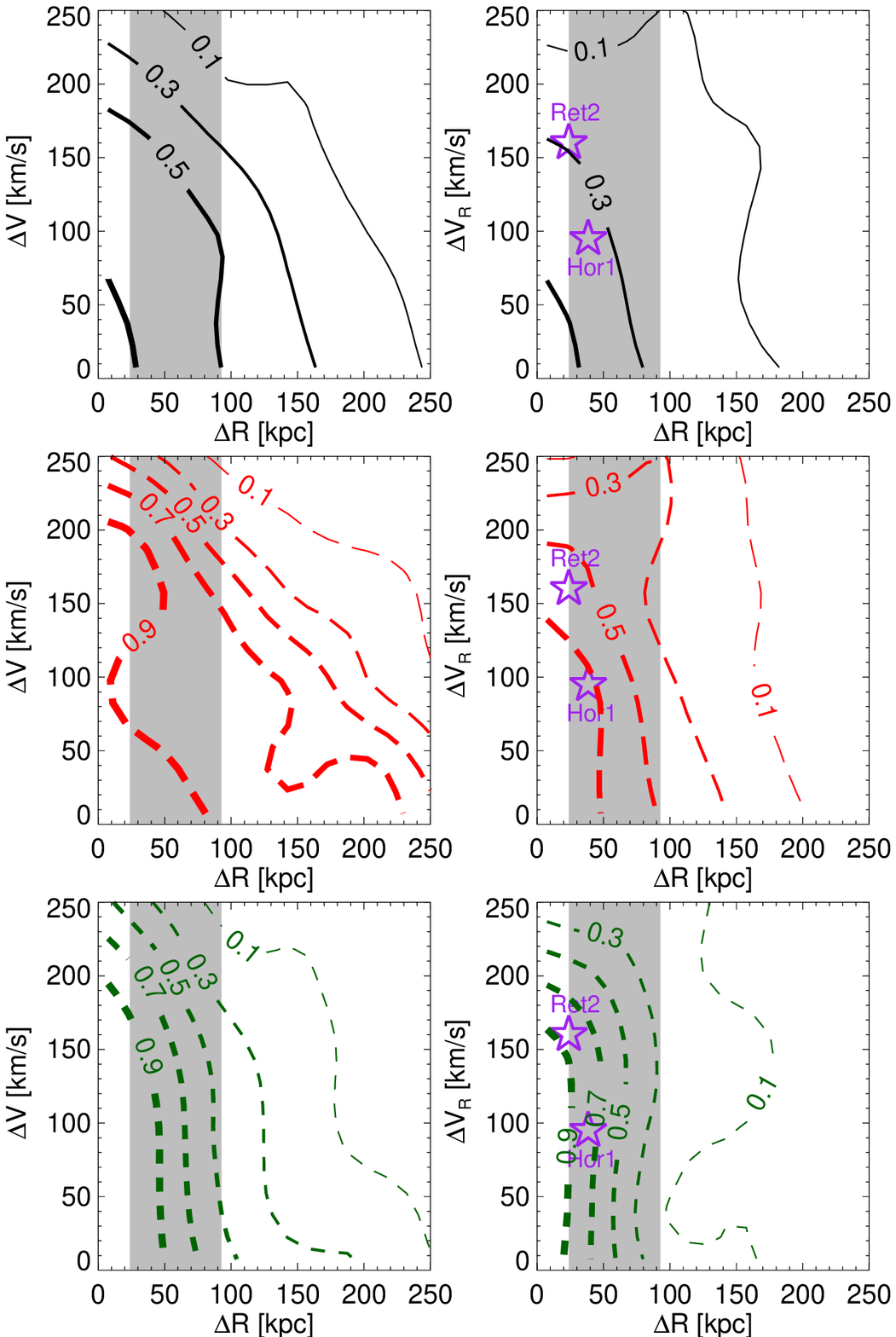}
   \caption{ \small The fraction of all MW/M31 satellites at $z=0$ that were a satellite of a surviving LMC-mass dwarf before infall onto the MW/M31 host as a function of 3D distance and velocity difference from the LMC. The contours levels give fractions of 0.1, 0.3, 0.5, 0.7 and 0.9 with increasing line thickness. The gray bands indicate the range of $\Delta R$ for eight candidate DES dwarfs (excluding Eri 2). The purple stars indicate the (spectroscopically confirmed) Ret 2 and Hor 1 dwarfs (\citealt{simon15}; \citealt{walker15}; \citealt{koposov15b}). All groups are shown in the top panels, recently accreted groups ($T_{\rm infall} < 2$ Gyr) are shown in the middle panels and the dynamical LMC-analogues are shown in the bottom panels.}
\label{fig:delta_prob_cont}
\end{figure}

The significance of infall time onto the MW/M31 host is further
illustrated in Fig. \ref{fig:delta_prob}. The red long-dashed and blue
dotted lines show the fraction of past members as a function of radial
and velocity difference for late ($T_{\rm infall} < 2$ Gyr) and early
($T_{\rm infall} > 5$ Gyr) accretion events, respectively. Groups
accreted a long time ago are now phase-mixed, whereas the probability
of being associated with a recently accreted LMC dwarf is strongly
related to the proximity in phase-space. We also show the probabilities when we only consider the dynamical analogues of the observed LMC with the green dashed line. These relations are very similar to the late accretion event subsample, likely because 2 (out of 3) of the dwarfs in this sample were accreted very recently (see Table \ref{tab:analogues}). Also, it is interesting that the average behavior of all recently-accreted LMC-analogues does not strongly differ from that of the dynamical analogues included here.

Combining both position and velocity information allows a much easier
distinction between previous members and the general satellite
population. In Fig. \ref{fig:delta_prob_cont} we show the joint probability distributions based on position and velocity. The top panels use the full sample of LMC-mass dwarfs, the middle panels only include late accretion events, and the bottom panels are for the dynamical LMC-dwarf analogues. The left-hand panels show that $\gtrsim 90$\% of dwarfs within 50 kpc and 50 km s$^{-1}$ of a
LMC-mass dwarf were likely once group members. For comparison, $\Delta
R = 23 \pm 2$ kpc and $\Delta V_{3D} = 128 \pm 32$ km s$^{-1}$
(\citealt{kalli13}) for the LMC-SMC pair.

We ensure that the results shown in Fig. \ref{fig:delta_prob} and Fig. \ref{fig:delta_prob_cont} are not significantly affected by the ``LMCs'' in our simulations that have very large pericenters. For example, we find that our results are unchanged if we restrict our sample to LMC-mass dwarfs with $R_{\rm peri} < 100$ kpc. Finally, one may expect that the probability of associations between satellites is affected by the Galactocentric distance of the LMC-mass dwarfs today, as there could be more ``interlopers'' at smaller radii from the MW/M31. However, we find no significant differences if we only include dwarfs inside (or outside) of $D=100$ kpc from the host center.

We can use these relations shown in Fig. \ref{fig:delta_prob} and Fig. \ref{fig:delta_prob_cont} to estimate the probability that the DES candidate dwarfs were once satellites of the LMC. The estimated probabilities are listed in Table \ref{tab:prob}. We also give the sum of these probabilities, which provides a rough estimate of the number of these dwarfs that are ``satellites of satellites''. Using only 3D coordinate information, we find that two of the DES dwarfs were once satellites of the LMC. If we assume that the LMC-group fell in very recently ($T_{\rm infall} < 2$ Gyr), or restrict our sample to dynamical analogues of the LMC, then this number rises to four. Fig. \ref{fig:delta_prob_cont} show that the inclusion of velocity information will enable a clearer distinction between members and non-members in the future.

\cite{simon15} and \cite{walker15} recently spectroscopically confirmed that the Ret 2 dwarf candidate is indeed a dwarf galaxy. Although this dwarf is in close proximity to the LMC ($\Delta R = 23.9$ kpc), the radial velocity measured by \cite{simon15} and \cite{walker15} for this dwarf is more disparate ($|\Delta V_R| = 160$ km s$^{-1}$)\footnote{Note that this is the difference in line-of-sight velocity between Ret 2 and the LMC in the Galactic rest frame.}. This dwarf is shown by a purple star in Fig. \ref{fig:delta_prob_cont}. With distance information alone we estimated $P_{\rm LMC \, \, sat} =0.38$, $P_{\rm LMC \, \, sat}(T_{\rm infall} < 2 \mathrm{Gyr})=0.65$ and $P_{\rm LMC \, \, sat}$ (dyn. analogues) $=0.77$, but this additional radial velocity information changes the probability of once being a satellite of the LMC to $P_{\rm LMC \, \, sat}=0.28$, $P_{\rm LMC \, \, sat}(T_{\rm infall} < 2 \mathrm{Gyr})=0.57$, $P_{\rm LMC \, \, sat}$ (dyn. analogues) $=0.84$ respectively. Finally, we also include the recent measurement of the radial velocity of Hor 1 by \cite{koposov15b}, to estimate probabilities of $P_{\rm LMC \, \, sat}=0.36$, $P_{\rm LMC \, \, sat}(T_{\rm infall} < 2 \mathrm{Gyr})=0.74$ and $P_{\rm LMC \, \, sat}$ (dyn. analogues) $=0.77$.

Note that we have not taken into account the presence of the SMC in the above analysis, and it is worth pointing out this potential caveat. It is beyond the scope of this work to quantify the affect of the LMC-SMC interaction on the orbits of lower-mass LMC satellites, but this could be a worthwhile avenue for further study.

\begin{table}
\centering
\renewcommand{\tabcolsep}{0.1cm}
\renewcommand{\arraystretch}{0.6}
\begin{tabular}{|l c c c c|}\hline
Name & $\Delta R$ & $P_{\rm LMC \,\, sat}$ & $P_{\rm LMC \, \,sat}$ & $P_{\rm LMC \, \,sat}$ \\
& [kpc] & & ($T_{\rm infall} < 2$ Gyr) &  \footnotesize{dyn. analogues}\\
\hline
Reticulum 2 & 23.9 & 0.38 & 0.65 & 0.77\\
Eridanus 2 & 337.4 & 0.02 & 0.01 & 0.04\\
Horologium 1 & 38.5 & 0.31 & 0.57 & 0.60\\
Pictoris 1 & 70.0 & 0.19 & 0.41 & 0.29\\
Phoenix 2 & 54.3 & 0.23  & 0.49 & 0.41\\
Indus 1 & 80.0 & 0.18 & 0.37 & 0.22\\
Grus 1 & 92.8 & 0.16 & 0.31 & 0.13\\
Eridanus 3 & 48.2 & 0.26 & 0.52  & 0.48\\
Tucana 2 & 36.7 & 0.32 & 0.58 & 0.62\\
\hline
\multicolumn{2}{r}{Total:} & 2.0 & 3.9 & 3.6\\
\hline
\end{tabular}
\caption[]{The nine candidate dwarf galaxies from \cite{koposov15}. We give the dwarf name, 3D distance from the LMC, and estimated probability of once being a satellite of the LMC based on this distance. The probabilities are computed using all LMC-mass satellites ($P_{\rm LMC \, \,sat}$), LMC-mass satellites with recent infall times ($P_{\rm LMC \, \,sat} [T_{\rm infall} < 2$ Gyr]) and LMC-mass satellites with similar dynamical constraints to the observed LMC ($P_{\rm LMC \, \,sat}$ [dyn. analogues]).}
\label{tab:prob}
\end{table}

\section{Conclusions}
We used the ELVIS simulation suite to study the surviving satellite
population of LMC-mass dwarfs accreted onto MW/M31 mass halos. A sample
of 25 LMC-mass ($M_{\rm peak} > 10^{11} M_\odot$) $z=0$ satellites of
MW/M31 hosts are selected, and we find the lower mass dwarfs that were
associated with these massive dwarfs before they fell into the MW/M31
hosts. Our selection is motivated by the recent discovery of nine
candidate dwarf galaxies in the vicinity of the LMC/SMC group. We also identify a subsample of 3 LMC-mass dwarfs with similar dynamical properties to the observed LMC-system, and we compare these with the overall sample. Our
main conclusions are summarized as follows:

\begin{itemize}

\item Recent, massive accretion events likely ``dragged in'' a
  significant number of MW/M31 dwarfs. Typically, 7\% of the surviving
  $z=0$ satellite population were once associated with surviving LMC-mass dwarfs,
  but this fraction can vary between 1\% and 25\% depending on the
  mass and infall time of the group central.

\item Groups of dwarfs quickly disperse in phase-space after infall
  onto MW/M31 mass hosts. We find that $z=0$ MW/M31 satellites that
  were once satellites of a surviving LMC dwarf can typically have large differences
  in velocity or configuration space relative to their group central
  if they fell into the MW/M31 host more than 5 Gyr ago.

\item The proximity of the candidate DES dwarfs to the LMC suggests
  that: (1) several were likely satellites of the LMC at some point in the
  past, and; (2) if they are genuine ``satellites of satellites'', then the LMC-group was likely accreted very recently ($\lesssim 2$ Gyr) for
  these dwarfs to retain such a close proximity in configuration space
  with the LMC. Distance information alone suggests that two to four of the newly
  discovered DES dwarfs were satellites of the LMC-group before
  infall.

\item The DES dwarfs that were/are satellites of the LMC could be
  prime candidates to study the affects of group pre-processing. If
  the LMC-group fell in very recently onto the MW, then the members
  may have spent a significant amount of time in this group before
  joining the MW. In future work, we plan to study the affects of
  group pre-processing in more detail.

\end{itemize}

\section*{Acknowledgments}
AJD is currently supported by NASA through Hubble Fellowship grant
HST-HF-51302.01, awarded by the Space Telescope Science Institute,
which is operated by the Association of Universities for Research in
Astronomy, Inc., for NASA, under contract NAS5-26555.
ARW gratefully acknowledges support from the Moore Center for Theoretical Cosmology and Physics at Caltech. We thank an anonymous referee for providing useful comments on the paper.

\label{lastpage}

\end{document}